\documentclass[twocolumn,english,aps,prb,floatfix,showpacs]{revtex4}
\usepackage[T1]{fontenc}
\usepackage[latin1]{inputenc}
\usepackage{graphicx}

\makeatletter

\usepackage{epsfig,amsmath,amssymb,color}
\usepackage{babel}
\makeatother
\begin{document}

\title{Order from Disorder in the Two-Dimensional Kondo-Necklace}

\author{Wolfram Brenig}

\affiliation{Technische Universit\"{a}t Braunschweig,
Mendelssohnstr. 3, 38106
Braunschweig, Germany}

\date{\today{}}

\begin{abstract}
We analyze the effects of site-dilution disorder on the thermodynamic
properties of the two-dimensional Kondo necklace using finite-temperature
stochastic series expansion. Results will be discussed for the dependence
on dilution concentration, temperature, and Kondo exchange-coupling
strength of the uniform susceptibility, the staggered structure factor,
and the Chakravarty-Halperin-Nelson ratio. Dilution is shown to induce
effective free-spin clusters in the gapped phase of the clean system
with a low-temperature Curie constant renormalized below 1/4. Furthermore,
dilution is demonstrated to generate antiferromagnetic order in the
quantum disordered phase of the clean system, i.e. order-from-disorder.
In turn, the quantum critical point of the clean system, separating
an antiferromagnetic from a paramagnetic dimerized state at a critical
Kondo exchange-coupling strength $J_{c}\approx1.4$ is suppressed.
Finally, speculations on a renormalized classical behavior in the
dilution induced ordered phase are stated.
\end{abstract}

\pacs{75.10.Jm, 05.70.Jk, 75.40.Cx, 75.40.Mg}

\maketitle
Quantum critical points (QCPs), i.e. zero temperature phase transitions
as a function of some control parameter are likely to be at the core
of unconventional finite temperature behavior of many novel materials
\cite{Chakravarty1989a,Chubukov1994a}. Quantum antiferromagnets (AFMs)
with an intrinsic spin dimerization and weak inter-dimer exchange
like K(Tl)CuCl$_{3}$ \cite{Tanaka1996a,Takatsu1997a} or BaCuSi$_{2}$O$_{6}$
\cite{Sasago1997a} are of particular interest here, since they allow
for switching between quantum disordered spin-gapped phases and states
with magnetic long-range order (LRO). QCPs have been induced in these
materials both, by tuning the inter-dimer exchange via pressure
\cite{Goto2004a,Oosawa2004a,Ruegg2004a},
as well as by applying external magnetic fields
\cite{Oosawa1999a,Nikuni2000a,Jaime2004a}.

Combining the physics of quantum critical spin systems with that of
disorder is an open issue. Site dilution with non-magnetic impurities
has been observed to induce LRO in the spin gapped phases of several
dimerized quantum AFMs \cite{Hase1993a,Azuma1997a,Oosawa2003a,Fujiwara2005a}.
In spin ladders and dimerized spin chains a picture of weakly interacting
'defect moments' has emerged with enhanced AFM correlations in the
vicinity of the non-magnetic sites
\cite{Martins1996a,Sigrist1996a,Miyazaki1997a,Laukamp1998a}.
In dimensions $D\geq2$ and at finite impurity concentration this
may trigger LRO in spin-gapped systems on bipartite lattices and remove
a QCP \cite{Wessel2001}. In the context of the cuprate parent compounds
\cite{Cheong1991a}, site dilution of the 2D Heisenberg AFM has been
shown to lead to a QCP at percolation where LRO is suppressed
\cite{Kato2000a,Sandvik2001a}.
In addition to site dilution several other forms of disorder are of
interest. Early on random exchange has been investigated in a variety
of 1D and 2D Heisenberg and Ising AFMs where it can lead to random
singlet fix-points \cite{Ma1979a,Fisher1994a,Fisher1995a} and Griffiths
phases \cite{Griffiths1969a}. Very recently 'dimer dilution' has
been studied in bilayer Heisenberg models yielding multicritical points
at percolation \cite{Sandvik2002a,Vajk2002a,Lin2006a}.

\begin{figure}
\begin{center}\includegraphics[%
  width=0.4\columnwidth,
  keepaspectratio]{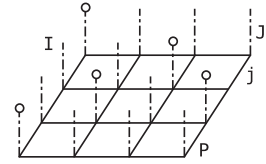}\end{center}
\vspace{-5mm}
\caption{\label{fig1}2D Kondo-necklace.
Spin-1/2 moments coupled by $j$ and $J$ and
located at vertices in 'conduction-plane' $P$
and on 'Kondo-sites' in plane $I$ at end of dash-dotted lines. '$\mathbf{\circ}$'
labels randomly diluted 'Kondo-sites' with no magnetic moment. }
\end{figure}

In this work we focus on random site dilution in the $SU\left(2\right)$
symmetric 2D spin-1/2 Kondo-necklace (SKN) which is shown in fig. 1
\begin{equation}
H_{SKN}=j\sum_{lm}\mathbf{S}_{Pl}\cdot\mathbf{S}_{Pm}+J\sum_{l}
\mathbf{S}_{Pl}\cdot\mathbf{S}_{Il}\label{eq:1}
\end{equation}
with $j\equiv1$ hereafter. This model can be viewed as the strong
coupling limit at large Coulomb correlations in the conduction electron
band of the Kondo-lattice model at half filling \cite{Brenig2006a}.
In turn, the two layers $P$ and $I$ in fig. \ref{fig1}, will be
loosely referred to as 'conduction-band' and 'Kondo-site' layer hereafter.
The clean limit of the 2D SKN has been studied at finite temperatures
recently \cite{Brenig2006a,Ling2006a}. The model exhibits a QCP between
AFM-LRO and a dimer phase at $J_{c}\approx1.4$. It shows temperature-scaling
and a Chakravarty-Halperin-Nelson ratio consistent with the $O\left(3\right)$
non-linear sigma model. The aim of this work is to shed light onto
the effect of randomly substituting non-magnetic ions into the Kondo-layer,
as shown in fig. \ref{fig1}. In that case the site index $l$ in
in the second sum of eqn. (\ref{eq:1}) runs only over those sites
which are \emph{occupied by a spin} in layer $I$.

Our analysis is based on the stochastic series expansion (SSE) with
loop-updates introduced by Sandvik and Syljuasen\cite{Sandvik1999a,sandvik02}.
Averaging over disorder configurations is performed by rapid thermalization
using the temperature-halving scheme proposed by Sandvik \cite{Sandvik2002b}.
We refer the reader to the latter three references for details. In
the following, all of the SSE results reported for a finite dilution
concentration comprise of an average over $1000$ disorder configurations
on a system of $N=24\times24\times2$ sites. The latter is justified
by the rather weak finite size dependence observed beyond this for
the clean system\cite{Brenig2006a}.

\begin{figure}[t]
\begin{center}\includegraphics[%
  width=0.8\columnwidth,
  keepaspectratio]{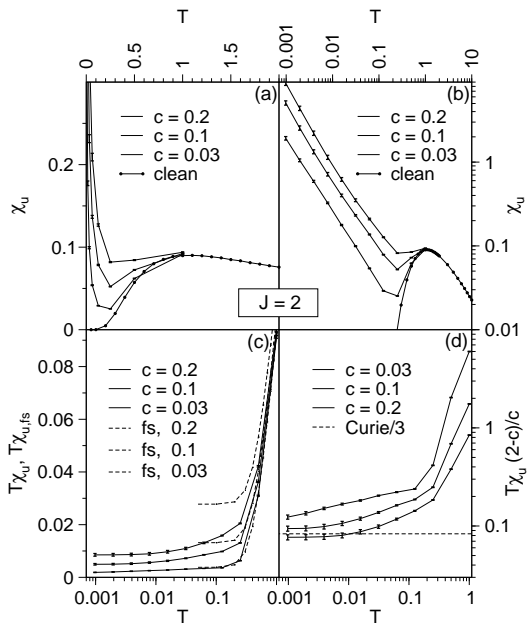}
\end{center}
\vspace{-3mm}
\caption{\label{fig2} (a), (b): Uniform susceptibility $\chi_{u}$ at 
$J=2>J_{c}$
vs. temperature for $c$=0.03, 0.1 and 0.2 and $1/1024\leq T\leq1$
and for $c=0$ and $0.05\leq T\leq1$ on a lin-lin (a) and a log-log
scale (b). (c): $T\chi_{u}$ product at $c\neq0$ as compared with
'free spin' approximation $T\chi_{u,fs}$ from eqn. (\ref{eq:3}), for
identical parameters as in (a). (d): $T\chi_{u}$ product at $c\neq0$
as compared to Curie behavior renormalized by 1/3. Statistical errors
in all panels either less than the solid-circle marker size or depicted
by error bars. Legends label plots from top to bottom.}
\end{figure}

To begin we discuss the uniform susceptibility
\begin{equation}
\chi_{u}=\left\langle m^{2}\right\rangle /T\,\,,\label{eq:2}
\end{equation}
where $T$ is the temperature and $m=\sum_{n=P+I,l}S_{nl}^{z}/N^{o}$
is the total spin $z$-component with $N^{o}$ being the total number
of sites \emph{occupied} by a spin. The defect concentration is 
$c=2N_{I}^{e}/N$
, where $N_{I}^{e}$ is the number of empty site in layer $I$, i.e.
$N^{o}=\left(2-c\right)N/2$. In fig. \ref{fig2} we summarize several
aspects of $\chi_{u}$ versus temperature at $J>J_{c}$ which refers
to the gapped phase of the clean system. The figure contrasts the
clean system for $0.05\leq T\leq1$ against three impurity concentrations
$c=0.03,\,0.1$, and $0.2$ for $1/1024\leq T\leq1$. First, fig.
\ref{fig2}(a) clearly demonstrates that the non-magnetic defects
induce low-energy magnetic density of states in the spin gap. Second,
the log-log plot of fig. \ref{fig2}(b) shows, that the susceptibility
of these states is similar to that of 'free' spins, i.e. it follows
almost a Curie law in the temperature range below the spin gap.

To further elucidate the free spin behavior, we start from the strong
coupling limit, i.e. $J\rightarrow\infty$, of the Kondo-necklace.
In that case and in the clean system the ground state is a product
of singlets on the dash-dotted bonds of fig. \ref{fig1}. Removing
a 'Kondo' spin leaves a free spin in the 'conduction' plane. Denoting
the uniform susceptibility per occupied site by $\chi_{u,fs}$, with
the index \emph{fs} referring to the free spins, one has
\begin{equation}
\chi_{u,fs}=\frac{2\left(1-c\right)}{\left(2-c\right)}
\chi_{u}\left(c=0\right)+\frac{c}{\left(2-c\right)}\,
\frac{1}{4T}\,\,\,.\label{eq:3}
\end{equation}
The first term is the contribution from the singlets and the second
term stems from the free spins. To assess the relevance of this free
spin picture - other than in the strong coupling limit - we may use
eqn. (\ref{eq:3}) with $\chi_{u}\left(c=0\right)$ taken from the
SSE at $c=0$ but for $J$ other than $\infty$. In fig. \ref{fig2}(c)
$T\chi_{u,fs}$ from this approximation is compared with $T\chi_{u}$
from eqn.(\ref{eq:2}) at $J=2$. First, at low temperatures $T\chi_{u}$
is smaller than $T\chi_{u,fs}$. Second, while $T\chi_{u,fs}$ saturates
on an energy scale set by the spin gap, $T\chi_{u}$ continues to
decrease as a function of $T$ and seem to saturate at far lower temperatures.
This is a clear indication of low-energy correlations which develop
between the impurity induced degrees of freedom. This also clarifies
the need for SSE results at far lower values of $T$ at finite disorder
as compared to the clean case.

\begin{figure}[t]
\begin{center}\includegraphics[%
  width=0.6\columnwidth,
  keepaspectratio]{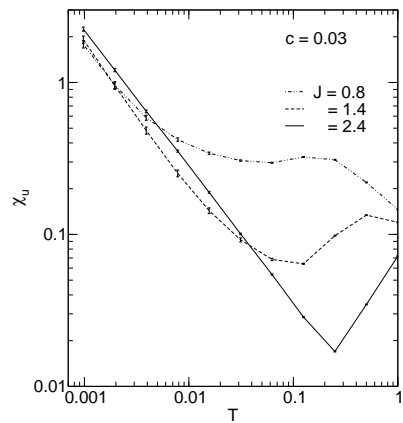}
\end{center}
\vspace{-3mm}
\caption{\label{fig3} Uniform susceptibility $\chi_{u}$ vs. temperature
at $c=0.03$ for $J=0.8$ (dashed-double-dotted), 1.4 (dashed), and
2.4 (solid), corresponding to the AFM-LRO, the critical and the gapped
regime at $c=0$. Statistical errors are depicted by vertical bars.}
\end{figure}

For site diluted spin-ladders Sigrist and Furusaki have proposed 
\cite{Sigrist1996a}
that below a crossover temperature, and due to long-range, higher
order exchange the free spins will form clusters of a size set by
the inverse temperature. These clusters lead to a Curie-type of 
susceptibility
per defect, however with a Curie constant reduced from $1/4$ to
$1/\left(3\cdot4\right)$. The argument of ref. \cite{Sigrist1996a}
is not restricted to ladders. To consider this, fig. \ref{fig2}(d)
shows the Curie contribution normalized to the defect concentration
on a log-log scale. For $c\geq0.1$ saturation of $T\chi_{u}$ can
be anticipated for $T<1/500$. Consistently with ref. \cite{Sigrist1996a}
$\lim_{T\rightarrow0}\left[T\chi_{u}\left(2-c\right)/c\right]<1/4$.
Moreover, while a renormalization by a factor of exactly $1/3$ can
not be read off from this figure, the reduction is very close to this
value. The kink in $T\chi_{u}$ observable at elevated temperatures
and small concentrations in this log-log plot is related to the opening
of the spin gap.

Fig. \ref{fig3} clarifies the difference between the gapped and the
ordered phase as well as the impact of finite size gaps. It contrast
to the gapped phase, site dilution of the LRO phase is not expected
to induce quasi free moments because of the AFM correlations between
all the occupied sites \cite{Sandvik1997a}. On finite systems however,
a size-dependend spin gap will remain
also in the AFM-LRO phase.
Disorder will eventually induce states
in this gap, leading to an increase of the uniform susceptibility,
however at a temperature scale much lower than and unrelated to $J$.
This is consistent with the results shown for small concentration
in fig. \ref{fig3}. In the gapped phase, at $J$=2.4 an almost straight-line
behavior can be observed on a log-log scale below an energy set by
the intrinsic spin gap. In the LRO phase at $J$=0.8 such behavior
can only be anticipated at temperatures at least two
orders of magnitude lower. 

\begin{figure}[t]
\begin{center}\includegraphics[%
  clip,
  width=0.95\columnwidth,
  keepaspectratio]{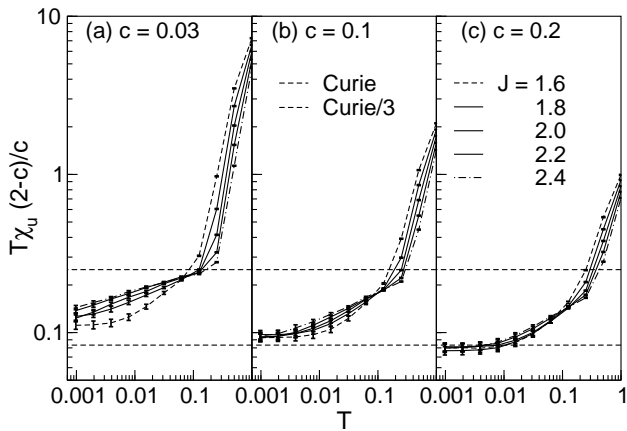}
\end{center}
\vspace{-3mm}
\caption{\label{fig4}$T\chi_{u}$ vs. $T$ at $c=0.03$ (a), 0.1 (b), and
0.2 (c) for various values of $J>J_{c}$. Dashed straight lines
correspond
to (renormalized) Curie constant. Legends label plots according to
order listed between dashed and dashed-dotted. Statistical errors
are depicted by vertical bars.}
\end{figure}

In fig. \ref{fig4} we summarize $T\chi_{u}$ versus $T$ for $c=0.03,\,0.1$,
and $0.2$ over a finite range of coupling constants $J$ corresponding
to the gapped regime of the clean system. First, while low-temperature
convergence to a constant $T\chi_{u}$ is observed for $c=0.1$ and
$0.2$, fig. \ref{fig4} allows for an approximate extrapolation to
$T=0$ at $c=0.03$ only. Second, it is tempting to claim that at
fixed $c$ the low temperature Curie constant depends only weakly
on $J$. Third, the low temperature Curie constant is less than $1/4$
and close to $1/\left(3\cdot4\right)$ for all cases depicted. However,
the figure suggests that the Curie constant is a \emph{decreasing}
function of $c$ and cannot be described by a single renormalization
factor at all $c$. This is consistent with the fact, that at $c=1$
the SKN is identical to the 2D AFM Heisenberg model, for which
$\lim_{T\rightarrow0}T\chi_{u}=0$.

\begin{figure}[t]
\begin{center}\includegraphics[%
  clip,
  width=0.95\columnwidth,
  keepaspectratio]{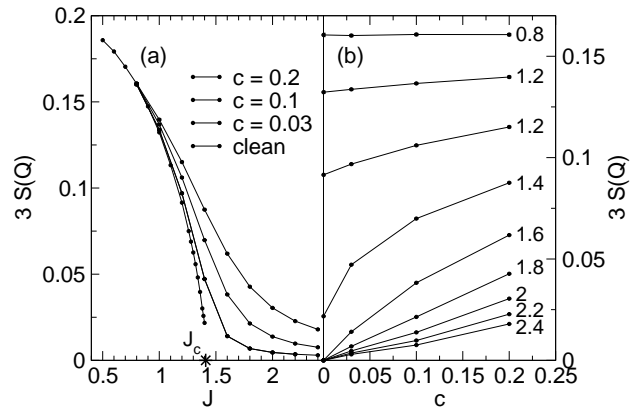}
\end{center}
\vspace{-3mm}
\caption{\label{fig5}Low-temperature staggered magnetization 
$3S\left(\mathbf{Q}\right)$
versus $J$ for various $c=0$, 0.03, 0.1, and 0.2 (a) and versus
$c$ for various $J=0.8\ldots\,2.4$ (b). Statistical errors are either
less than the solid-circle marker size or are depicted by vertical
bars. Legends label plots from top to bottom.}
\end{figure}

Now we turn to the longitudinal staggered structure factor, i.e. the
order parameter for AFM-LRO
\begin{equation}
S\left(\mathbf{Q}\right)=\left\langle 
\left(m_{\mathbf{Q}}^{z}\right)^{2}\right\rangle 
\,\,,\label{eq:4}
\end{equation}
 where $m_{\mathbf{Q}}^{z}=\sum_{n=P+I,l}S_{nl}^{z}
\exp\left(i\mathbf{Q}\cdot\mathbf{r}_{l}\right)/N^{o}$
is the staggered magnetization with $\mathbf{Q}=\left(\pi,\pi,\pi\right)$.
All results discussed correspond to values of $T$ such that low temperature
saturation of $S\left(\mathbf{Q}\right)$ has been reached. Fig. 
\ref{fig5}(a)
shows the squared staggered moment $M_{\mathbf{Q}}^{2}=3S
\left(\mathbf{Q}\right)$
versus $J$ for the clean case and for $c$=0.03, 0.1 and 0.2. In
the clean case $M_{\mathbf{Q}}^{2}$ is finite below the critical value
of $J=J_{c}$ and, apart from finite size effects, drops to zero
for $J>J_{c}$. I.e. $J_{c}$ corresponds to the QCP \cite{Brenig2006a}
with AFM-LRO for $J<J_{c}$ and a spin gapped state for $J>J_{c}$.
Generating SSE data of smaller value of $M_{\mathbf{Q}}^{2}$ for
$J\sim J_{c}$ at $c=0$ requires 'fine-tuning' of $J$ which we refrain
from.

The main result of fig. \ref{fig5}(a) is contained in the change
of $M_{\mathbf{Q}}^{2}$ versus $J$ upon doping. First, for all 
concentrations
investigated the QCP disappears. Second, a finite staggered moment
can be observed at all values of $J$ and in particular also in the
formerly gapped phase of the clean system. This implies that doping
by non magnetic impurities induces AFM-LRO in the quantum disorder
phase of the SKN, i.e. \emph{order-from-disorder}. It is tempting
to speculate, that this behavior is true for all $c$.

We may also discuss these results from a different point of view,
i.e. by considering $M_{\mathbf{Q}}^{2}$ versus the impurity concentration
for various exchange coupling constants, as in fig. \ref{fig5}(b).
This shows $M_{\mathbf{Q}}^{2}$ to increase at small $c$, both below
and above $J_{c}$. At $J_{c}$ the induced AFM order is most sensitive
to the impurity concentration. For $c\rightarrow1$ all curves are
expected to join at a single value corresponding to the staggered
magnetization of the planar Heisenberg AFM. For $c=0$ and $J\geq1.6>J_{c}$,
$M_{\mathbf{Q}}^{2}$ has been forced to zero in fig. \ref{fig5}(b).
This neglects finite size effects, namely that $M_{\mathbf{Q}}^{2}\neq0$,
albeit small, even for $c=0$ since $N=24$ is finite. In principle
this value of $M_{\mathbf{Q}}^{2}\left(J>J_{c},c=0\right)$ can be
read off by extrapolating to $c\rightarrow0$ for each $J>J_{c}$
in fig. \ref{fig5}(b).

Finally we discuss the Chakravarty-Halperin-Nelson ratio
\begin{equation}
R=\frac{S\left(\mathbf{Q}\right)}{T\chi\left(\mathbf{Q}\right)}\,.
\label{eq:5}\end{equation}
which, in addition to $S\left(\mathbf{Q}\right)$ includes information
on the longitudinal staggered susceptibility

\begin{equation}
\chi\left(\mathbf{Q}\right)=\int_{0}^{1/T}d\tau\left\langle 
m_{\mathbf{Q}}^{z}\left(\tau\right)m_{\mathbf{Q}}^{z}\right\rangle
\label{eq:6}\end{equation}

\begin{figure}[t]
\begin{center}\includegraphics[%
  width=0.8\columnwidth]{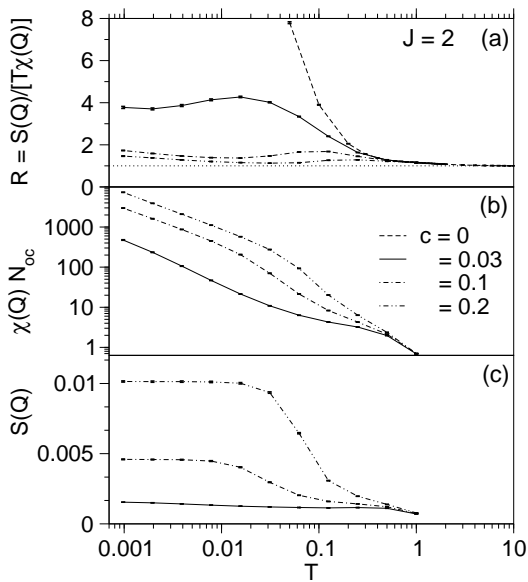}
\end{center}
\caption{\label{fig6}(a): Chakravarty-Halperin-Nelson ratio $R$ of the total
staggered structure factor and susceptibility vs. temperature at $J=2>J_{c}$
for $c=0$, 0.03, 0.1, and 0.2, corresponding to dashed, solid, dashed-dotted
and dashed-double-dotted. Result for $c=0$ in panel (a) from ref.
\cite{Brenig2006a}. Temperatures are $1/1024\leq T\leq1$ ($0.05\leq T\leq10$)
for $c\neq0$ ($c=0$). Dotted line: unity. (b) and (c): denominator $\chi\left(\mathbf{Q}\right)$
and numerator $S\left(\mathbf{Q}\right)$ from eqn. (\ref{eq:5})
for $R$ vs. temperature with $c$ as in fig. \ref{fig6} (a). Size
of statistical errors is given by vertical bars in all panels.}
\end{figure}

 This ratio relates our analysis to that of the AFM non-linear
$\sigma$-model
(NL$\sigma$M) \cite{Chakravarty1989a,Chubukov1994a,Sokol1994a}.
Regarding clean systems with a QCP between AFM-LRO and dimerization,
this ratio has been studied in detail for the bilayer Heisenberg AFM
\cite{Shevchenko2000a} and for the SKN \cite{Brenig2006a}. In agreement
with the NL$\sigma$M it was found that $R=1$ in the (classical)
high-$T$ regime, as well as in AMF-LRO (renormalized classical) regime
for $T\rightarrow0$. At the QCP the NL$\sigma$M requires
$R\left(T\rightarrow0\right)\simeq1.09$,
which is also consistent with SSE at $J=J_{c}$. While - to our knowledge
- rigorous results are absent for a disorder induced AFM-LRO phase,
as for $J>J_{c}$, it is yet tempting to speculate that renormalized
classical behavior will re-emerge. Since in the gaped phase of the
clean system $R$ diverges as $T^{-1}$ this will lead to drastic
variations in $R$ versus $c$ for $J>J_{c}$. Indeed this is observed
in fig. \ref{fig6}(a), where $R$ is shown versus $T$ as $c$ varies
from 0 to 0.2. Obviously $R$ decreases markedly with increasing $c$
and almost reaches $1$ for $c\geq0.1$ and $T\sim0.01$. As for $c=0$
the zero temperature limit of $R$ will be very sensitive to finite
size gaps\cite{Shevchenko2000a,Brenig2006a}. This may cause the low-$T$
increase seen in this figure, i.e. we speculate that for larger systems
$R=1$ as $T\rightarrow0$ for all $c\neq0$. This has to clarified
by future analysis. For completeness, fig. \ref{fig6}(b) and (c)
relate the cancellation of the factor $T$ in the denominator of $R$
at $c\neq0$ to the temperature dependence
of $\chi\left(\mathbf{Q}\right)$.

\emph{Acknowledgements} Part of this work has been supported by
DFG grant No. BR 1084/4-1. Kind hospitality of the Kavli
Institute for Theoretical Physics, Santa Barbara
in the early stages of this project is gratefully
acknowledged where parts of this work have been
supported by NSF grant No. PHY99-07949.


\begin{thebibliography}{38}
\expandafter\ifx\csname natexlab\endcsname\relax\def\natexlab#1{#1}\fi
\expandafter\ifx\csname bibnamefont\endcsname\relax
  \def\bibnamefont#1{#1}\fi
\expandafter\ifx\csname bibfnamefont\endcsname\relax
  \def\bibfnamefont#1{#1}\fi
\expandafter\ifx\csname citenamefont\endcsname\relax
  \def\citenamefont#1{#1}\fi
\expandafter\ifx\csname url\endcsname\relax
  \def\url#1{\texttt{#1}}\fi
\expandafter\ifx\csname urlprefix\endcsname\relax\def\urlprefix{URL }\fi
\providecommand{\bibinfo}[2]{#2}
\providecommand{\eprint}[2][]{\url{#2}}

\bibitem[{\citenamefont{Chakravarty et~al.}(1989)\citenamefont{Chakravarty,
  Halperin, and Nelson}}]{Chakravarty1989a}
\bibinfo{author}{\bibfnamefont{S.}~\bibnamefont{Chakravarty}},
  \bibinfo{author}{\bibfnamefont{B.~I.} \bibnamefont{Halperin}},
  \bibnamefont{and} \bibinfo{author}{\bibfnamefont{D.~R.}
  \bibnamefont{Nelson}}, \bibinfo{journal}{Phys. Rev. B}
  \textbf{\bibinfo{volume}{39}}, \bibinfo{pages}{2344} (\bibinfo{year}{1989}).

\bibitem[{\citenamefont{Chubukov et~al.}(1994)\citenamefont{Chubukov, Sachdev,
  and Ye}}]{Chubukov1994a}
\bibinfo{author}{\bibfnamefont{A.~V.} \bibnamefont{Chubukov}},
  \bibinfo{author}{\bibfnamefont{S.}~\bibnamefont{Sachdev}}, \bibnamefont{and}
  \bibinfo{author}{\bibfnamefont{J.}~\bibnamefont{Ye}}, \bibinfo{journal}{Phys.
  Rev. B} \textbf{\bibinfo{volume}{49}}, \bibinfo{pages}{11919}
  (\bibinfo{year}{1994}).

\bibitem[{\citenamefont{Tanaka et~al.}(1996)\citenamefont{Tanaka, Takatsu,
  Shiramura, and Ono}}]{Tanaka1996a}
\bibinfo{author}{\bibfnamefont{H.}~\bibnamefont{Tanaka}},
  \bibinfo{author}{\bibfnamefont{K.}~\bibnamefont{Takatsu}},
  \bibinfo{author}{\bibfnamefont{W.}~\bibnamefont{Shiramura}},
  \bibnamefont{and} \bibinfo{author}{\bibfnamefont{T.}~\bibnamefont{Ono}},
  \bibinfo{journal}{J. Phys. Soc. Jpn.} \textbf{\bibinfo{volume}{65}},
  \bibinfo{pages}{1945} (\bibinfo{year}{1996}).

\bibitem[{\citenamefont{Takatsu et~al.}(1997)\citenamefont{Takatsu, Shiramura,
  and Tanaka}}]{Takatsu1997a}
\bibinfo{author}{\bibfnamefont{K.}~\bibnamefont{Takatsu}},
  \bibinfo{author}{\bibfnamefont{W.}~\bibnamefont{Shiramura}},
  \bibnamefont{and} \bibinfo{author}{\bibfnamefont{H.}~\bibnamefont{Tanaka}},
  \bibinfo{journal}{J. Phys. Soc. Jpn.} \textbf{\bibinfo{volume}{66}},
  \bibinfo{pages}{1611} (\bibinfo{year}{1997}).

\bibitem[{\citenamefont{Sasago et~al.}(1997)\citenamefont{Sasago, Zheludev, and
  Shirane}}]{Sasago1997a}
\bibinfo{author}{\bibfnamefont{Y.}~\bibnamefont{Sasago}},
  \bibinfo{author}{\bibfnamefont{K.~U.~A.} \bibnamefont{Zheludev}},
  \bibnamefont{and} \bibinfo{author}{\bibfnamefont{G.}~\bibnamefont{Shirane}},
  \bibinfo{journal}{Phys. Rev. B} \textbf{\bibinfo{volume}{55}},
  \bibinfo{pages}{8357} (\bibinfo{year}{1997}).

\bibitem[{\citenamefont{Goto et~al.}(2004)\citenamefont{Goto, Fujisawa, Ono,
  Tanaka, and Uwatoko}}]{Goto2004a}
\bibinfo{author}{\bibfnamefont{K.}~\bibnamefont{Goto}},
  \bibinfo{author}{\bibfnamefont{M.}~\bibnamefont{Fujisawa}},
  \bibinfo{author}{\bibnamefont{Ono}}, \bibinfo{author}{\bibnamefont{Tanaka}},
  \bibnamefont{and} \bibinfo{author}{\bibfnamefont{Y.}~\bibnamefont{Uwatoko}},
  \bibinfo{journal}{J. Phys. Soc. Jpn.} \textbf{\bibinfo{volume}{73}},
  \bibinfo{pages}{3254} (\bibinfo{year}{2004}).

\bibitem[{\citenamefont{Oosawa et~al.}(2004)\citenamefont{Oosawa, Kakurai,
  Osakabe, Nakamura, Takeda, and Tanaka}}]{Oosawa2004a}
\bibinfo{author}{\bibfnamefont{A.}~\bibnamefont{Oosawa}},
  \bibinfo{author}{\bibfnamefont{K.}~\bibnamefont{Kakurai}},
  \bibinfo{author}{\bibfnamefont{T.}~\bibnamefont{Osakabe}},
  \bibinfo{author}{\bibfnamefont{M.}~\bibnamefont{Nakamura}},
  \bibinfo{author}{\bibfnamefont{M.}~\bibnamefont{Takeda}}, \bibnamefont{and}
  \bibinfo{author}{\bibfnamefont{H.}~\bibnamefont{Tanaka}},
  \bibinfo{journal}{J. Phys. Soc. Jpn.} \textbf{\bibinfo{volume}{73}},
  \bibinfo{pages}{1446} (\bibinfo{year}{2004}).

\bibitem[{\citenamefont{R\"{u}egg et~al.}(2004)\citenamefont{R\"{u}egg, Furrer,
  Sheptyakov, Str\"{a}ssle, Kr\"{a}mer, G\"{u}del, and
  M\'{e}l\'{e}si}}]{Ruegg2004a}
\bibinfo{author}{\bibfnamefont{C.}~\bibnamefont{R\"{u}egg}},
  \bibinfo{author}{\bibfnamefont{A.}~\bibnamefont{Furrer}},
  \bibinfo{author}{\bibfnamefont{D.}~\bibnamefont{Sheptyakov}},
  \bibinfo{author}{\bibfnamefont{T.}~\bibnamefont{Str\"{a}ssle}},
  \bibinfo{author}{\bibfnamefont{K.~W.} \bibnamefont{Kr\"{a}mer}},
  \bibinfo{author}{\bibfnamefont{H.-U.} \bibnamefont{G\"{u}del}},
  \bibnamefont{and}
  \bibinfo{author}{\bibfnamefont{L.}~\bibnamefont{M\'{e}l\'{e}si}},
  \bibinfo{journal}{Phys. Rev. Lett.} \textbf{\bibinfo{volume}{93}},
  \bibinfo{pages}{257201} (\bibinfo{year}{2004}).

\bibitem[{\citenamefont{Oosawa et~al.}(1999)\citenamefont{Oosawa, Ishii, and
  Tanaka}}]{Oosawa1999a}
\bibinfo{author}{\bibfnamefont{A.}~\bibnamefont{Oosawa}},
  \bibinfo{author}{\bibfnamefont{M.}~\bibnamefont{Ishii}}, \bibnamefont{and}
  \bibinfo{author}{\bibfnamefont{H.}~\bibnamefont{Tanaka}},
  \bibinfo{journal}{J. Phys.: Condens. Matter} \textbf{\bibinfo{volume}{11}},
  \bibinfo{pages}{265} (\bibinfo{year}{1999}).

\bibitem[{\citenamefont{Nikuni et~al.}(2000)\citenamefont{Nikuni, Oshikawa,
  Oosawa, and Tanaka}}]{Nikuni2000a}
\bibinfo{author}{\bibfnamefont{T.}~\bibnamefont{Nikuni}},
  \bibinfo{author}{\bibfnamefont{M.}~\bibnamefont{Oshikawa}},
  \bibinfo{author}{\bibfnamefont{A.}~\bibnamefont{Oosawa}}, \bibnamefont{and}
  \bibinfo{author}{\bibfnamefont{H.}~\bibnamefont{Tanaka}},
  \bibinfo{journal}{Phys. Rev. Lett.} \textbf{\bibinfo{volume}{84}},
  \bibinfo{pages}{5868} (\bibinfo{year}{2000}).

\bibitem[{\citenamefont{Jaime et~al.}(2004)\citenamefont{Jaime, Correa,
  Harrison, Batista, Kawashima, Kazuma, Jorge, Stern, Heinmaa, Zvyagin
  et~al.}}]{Jaime2004a}
\bibinfo{author}{\bibfnamefont{M.}~\bibnamefont{Jaime}},
  \bibinfo{author}{\bibfnamefont{V.~F.} \bibnamefont{Correa}},
  \bibinfo{author}{\bibfnamefont{N.}~\bibnamefont{Harrison}},
  \bibinfo{author}{\bibfnamefont{C.~D.} \bibnamefont{Batista}},
  \bibinfo{author}{\bibfnamefont{N.}~\bibnamefont{Kawashima}},
  \bibinfo{author}{\bibfnamefont{Y.}~\bibnamefont{Kazuma}},
  \bibinfo{author}{\bibfnamefont{G.~A.} \bibnamefont{Jorge}},
  \bibinfo{author}{\bibfnamefont{R.}~\bibnamefont{Stern}},
  \bibinfo{author}{\bibfnamefont{I.}~\bibnamefont{Heinmaa}},
  \bibinfo{author}{\bibfnamefont{S.~A.} \bibnamefont{Zvyagin}},
  \bibnamefont{et~al.}, \bibinfo{journal}{Phys. Rev. Lett.}
  \textbf{\bibinfo{volume}{93}}, \bibinfo{pages}{087203}
  (\bibinfo{year}{2004}).

\bibitem[{\citenamefont{Hase et~al.}(1993)\citenamefont{Hase, Terasaki, Sasago,
  Uchinokura, and Obara}}]{Hase1993a}
\bibinfo{author}{\bibfnamefont{M.~.} \bibnamefont{Hase}},
  \bibinfo{author}{\bibfnamefont{I.}~\bibnamefont{Terasaki}},
  \bibinfo{author}{\bibfnamefont{Y.}~\bibnamefont{Sasago}},
  \bibinfo{author}{\bibfnamefont{K.}~\bibnamefont{Uchinokura}},
  \bibnamefont{and} \bibinfo{author}{\bibfnamefont{H.}~\bibnamefont{Obara}},
  \bibinfo{journal}{Phys. Rev. Lett.} \textbf{\bibinfo{volume}{71}},
  \bibinfo{pages}{4059} (\bibinfo{year}{1993}).

\bibitem[{\citenamefont{Azuma et~al.}(1997)\citenamefont{Azuma, Fujishiro,
  Takano, Nohara, and Takagi}}]{Azuma1997a}
\bibinfo{author}{\bibfnamefont{M.}~\bibnamefont{Azuma}},
  \bibinfo{author}{\bibfnamefont{Y.}~\bibnamefont{Fujishiro}},
  \bibinfo{author}{\bibfnamefont{M.}~\bibnamefont{Takano}},
  \bibinfo{author}{\bibfnamefont{M.}~\bibnamefont{Nohara}}, \bibnamefont{and}
  \bibinfo{author}{\bibfnamefont{H.}~\bibnamefont{Takagi}},
  \bibinfo{journal}{Phys. Rev. B} \textbf{\bibinfo{volume}{55}},
  \bibinfo{pages}{R8658} (\bibinfo{year}{1997}).

\bibitem[{\citenamefont{Oosawa et~al.}(2003)\citenamefont{Oosawa, Fujisawa,
  Kakurai, and Tanaka}}]{Oosawa2003a}
\bibinfo{author}{\bibfnamefont{A.}~\bibnamefont{Oosawa}},
  \bibinfo{author}{\bibfnamefont{M.}~\bibnamefont{Fujisawa}},
  \bibinfo{author}{\bibfnamefont{K.}~\bibnamefont{Kakurai}}, \bibnamefont{and}
  \bibinfo{author}{\bibfnamefont{H.}~\bibnamefont{Tanaka}},
  \bibinfo{journal}{Phys. Rev. B} \textbf{\bibinfo{volume}{67}},
  \bibinfo{pages}{184424} (\bibinfo{year}{2003}).

\bibitem[{\citenamefont{Fujiwara et~al.}(2005)\citenamefont{Fujiwara, Shindo,
  and Tanaka}}]{Fujiwara2005a}
\bibinfo{author}{\bibfnamefont{H.}~\bibnamefont{Fujiwara}},
  \bibinfo{author}{\bibfnamefont{Y.}~\bibnamefont{Shindo}}, \bibnamefont{and}
  \bibinfo{author}{\bibfnamefont{H.}~\bibnamefont{Tanaka}},
  \bibinfo{journal}{Prog. Theor. Phys. Suppl.} \textbf{\bibinfo{volume}{159}},
  \bibinfo{pages}{392} (\bibinfo{year}{2005}).

\bibitem[{\citenamefont{Martins et~al.}(1996)\citenamefont{Martins, Dagotto,
  and Riera}}]{Martins1996a}
\bibinfo{author}{\bibfnamefont{G.~B.} \bibnamefont{Martins}},
  \bibinfo{author}{\bibfnamefont{E.}~\bibnamefont{Dagotto}}, \bibnamefont{and}
  \bibinfo{author}{\bibfnamefont{J.~A.} \bibnamefont{Riera}},
  \bibinfo{journal}{Phys. Rev. B} \textbf{\bibinfo{volume}{54}},
  \bibinfo{pages}{16032} (\bibinfo{year}{1996}).

\bibitem[{\citenamefont{Sigrist and Furusaki}(1996)}]{Sigrist1996a}
\bibinfo{author}{\bibfnamefont{M.}~\bibnamefont{Sigrist}} \bibnamefont{and}
  \bibinfo{author}{\bibfnamefont{A.}~\bibnamefont{Furusaki}},
  \bibinfo{journal}{J. Phys. Soc. Jpn.} \textbf{\bibinfo{volume}{65}},
  \bibinfo{pages}{2385} (\bibinfo{year}{1996}).

\bibitem[{\citenamefont{Miyazaki et~al.}(1997)\citenamefont{Miyazaki, Troyer,
  Ogata, Ueda, and Yoshioka}}]{Miyazaki1997a}
\bibinfo{author}{\bibfnamefont{T.}~\bibnamefont{Miyazaki}},
  \bibinfo{author}{\bibfnamefont{M.}~\bibnamefont{Troyer}},
  \bibinfo{author}{\bibfnamefont{M.}~\bibnamefont{Ogata}},
  \bibinfo{author}{\bibfnamefont{K.}~\bibnamefont{Ueda}}, \bibnamefont{and}
  \bibinfo{author}{\bibfnamefont{D.}~\bibnamefont{Yoshioka}},
  \bibinfo{journal}{J. Phys. Soc. Jpn.} \textbf{\bibinfo{volume}{66}},
  \bibinfo{pages}{2580} (\bibinfo{year}{1997}).

\bibitem[{\citenamefont{Laukamp et~al.}(1998)\citenamefont{Laukamp, Martins,
  Gazza, Malvezzi, Dagotto, Hansen, L\'opez, and Riera}}]{Laukamp1998a}
\bibinfo{author}{\bibfnamefont{M.}~\bibnamefont{Laukamp}},
  \bibinfo{author}{\bibfnamefont{G.~B.} \bibnamefont{Martins}},
  \bibinfo{author}{\bibfnamefont{C.}~\bibnamefont{Gazza}},
  \bibinfo{author}{\bibfnamefont{A.~L.} \bibnamefont{Malvezzi}},
  \bibinfo{author}{\bibfnamefont{E.}~\bibnamefont{Dagotto}},
  \bibinfo{author}{\bibfnamefont{P.~M.} \bibnamefont{Hansen}},
  \bibinfo{author}{\bibfnamefont{A.~C.} \bibnamefont{L\'opez}},
  \bibnamefont{and} \bibinfo{author}{\bibfnamefont{J.}~\bibnamefont{Riera}},
  \bibinfo{journal}{Phys. Rev. B} \textbf{\bibinfo{volume}{57}},
  \bibinfo{pages}{10755} (\bibinfo{year}{1998}).

\bibitem[{\citenamefont{Wessel et~al.}(2001)\citenamefont{Wessel, Normand,
  Sigrist, and Haas}}]{Wessel2001}
\bibinfo{author}{\bibfnamefont{S.}~\bibnamefont{Wessel}},
  \bibinfo{author}{\bibfnamefont{B.}~\bibnamefont{Normand}},
  \bibinfo{author}{\bibfnamefont{M.}~\bibnamefont{Sigrist}}, \bibnamefont{and}
  \bibinfo{author}{\bibfnamefont{S.}~\bibnamefont{Haas}},
  \bibinfo{journal}{Phys. Rev. Lett.} \textbf{\bibinfo{volume}{86}},
  \bibinfo{pages}{1086} (\bibinfo{year}{2001}).

\bibitem[{\citenamefont{Cheong et~al.}(1991)\citenamefont{Cheong, Cooper, Rupp,
  Batlogg, Thompson, and Fisk}}]{Cheong1991a}
\bibinfo{author}{\bibfnamefont{S.~W.} \bibnamefont{Cheong}},
  \bibinfo{author}{\bibfnamefont{A.~S.} \bibnamefont{Cooper}},
  \bibinfo{author}{\bibfnamefont{L.~W.} \bibnamefont{Rupp}},
  \bibinfo{author}{\bibfnamefont{B.}~\bibnamefont{Batlogg}},
  \bibinfo{author}{\bibfnamefont{J.~D.} \bibnamefont{Thompson}},
  \bibnamefont{and} \bibinfo{author}{\bibfnamefont{Z.}~\bibnamefont{Fisk}},
  \bibinfo{journal}{Phys. Rev. Lett.} \textbf{\bibinfo{volume}{44}},
  \bibinfo{pages}{9739} (\bibinfo{year}{1991}).

\bibitem[{\citenamefont{Kato et~al.}(2000)\citenamefont{Kato, Todo, Harada,
  Kawashima, Miyashita, and Takayama}}]{Kato2000a}
\bibinfo{author}{\bibfnamefont{K.}~\bibnamefont{Kato}},
  \bibinfo{author}{\bibfnamefont{S.}~\bibnamefont{Todo}},
  \bibinfo{author}{\bibfnamefont{K.}~\bibnamefont{Harada}},
  \bibinfo{author}{\bibfnamefont{N.}~\bibnamefont{Kawashima}},
  \bibinfo{author}{\bibfnamefont{S.}~\bibnamefont{Miyashita}},
  \bibnamefont{and} \bibinfo{author}{\bibfnamefont{H.}~\bibnamefont{Takayama}},
  \bibinfo{journal}{Phys. Rev. Lett.} \textbf{\bibinfo{volume}{84}},
  \bibinfo{pages}{4204} (\bibinfo{year}{2000}).

\bibitem[{\citenamefont{Sandvik}(2001)}]{Sandvik2001a}
\bibinfo{author}{\bibfnamefont{A.~W.} \bibnamefont{Sandvik}},
  \bibinfo{journal}{Phys. Rev. Lett.} \textbf{\bibinfo{volume}{86}},
  \bibinfo{pages}{3209} (\bibinfo{year}{2001}).

\bibitem[{\citenamefont{Ma et~al.}(1979)\citenamefont{Ma, Dasgupta, and
  Hu}}]{Ma1979a}
\bibinfo{author}{\bibfnamefont{S.~K.} \bibnamefont{Ma}},
  \bibinfo{author}{\bibfnamefont{C.}~\bibnamefont{Dasgupta}}, \bibnamefont{and}
  \bibinfo{author}{\bibfnamefont{C.}~\bibnamefont{Hu}}, \bibinfo{journal}{Phys.
  Rev. Lett.} \textbf{\bibinfo{volume}{43}}, \bibinfo{pages}{1434}
  (\bibinfo{year}{1979}).

\bibitem[{\citenamefont{Fisher}(1994)}]{Fisher1994a}
\bibinfo{author}{\bibfnamefont{D.~S.} \bibnamefont{Fisher}},
  \bibinfo{journal}{Phys. Rev. B} \textbf{\bibinfo{volume}{50}},
  \bibinfo{pages}{3799} (\bibinfo{year}{1994}).

\bibitem[{\citenamefont{Fisher}(1995)}]{Fisher1995a}
\bibinfo{author}{\bibfnamefont{D.~S.} \bibnamefont{Fisher}},
  \bibinfo{journal}{Phys. Rev. B} \textbf{\bibinfo{volume}{51}},
  \bibinfo{pages}{6411} (\bibinfo{year}{1995}).

\bibitem[{\citenamefont{Griffiths}(1969)}]{Griffiths1969a}
\bibinfo{author}{\bibfnamefont{R.~B.} \bibnamefont{Griffiths}},
  \bibinfo{journal}{Phys. Rev. Lett.} \textbf{\bibinfo{volume}{23}},
  \bibinfo{pages}{17} (\bibinfo{year}{1969}).

\bibitem[{\citenamefont{Sandvik}(2002{\natexlab{a}})}]{Sandvik2002a}
\bibinfo{author}{\bibfnamefont{A.~W.} \bibnamefont{Sandvik}},
  \bibinfo{journal}{Phys. Rev. Lett.} \textbf{\bibinfo{volume}{89}},
  \bibinfo{pages}{177201} (\bibinfo{year}{2002}{\natexlab{a}}).

\bibitem[{\citenamefont{Vajk and Greven}(2002)}]{Vajk2002a}
\bibinfo{author}{\bibfnamefont{O.~P.} \bibnamefont{Vajk}} \bibnamefont{and}
  \bibinfo{author}{\bibfnamefont{M.}~\bibnamefont{Greven}},
  \bibinfo{journal}{Phys. Rev. Lett.} \textbf{\bibinfo{volume}{89}},
  \bibinfo{pages}{177202} (\bibinfo{year}{2002}).

\bibitem[{\citenamefont{Lin et~al.}()\citenamefont{Lin, Rieger, Laflorencie,
  and Igloi}}]{Lin2006a}
\bibinfo{author}{\bibfnamefont{Y.~C.} \bibnamefont{Lin}},
  \bibinfo{author}{\bibfnamefont{H.}~\bibnamefont{Rieger}},
  \bibinfo{author}{\bibfnamefont{N.}~\bibnamefont{Laflorencie}},
  \bibnamefont{and} \bibinfo{author}{\bibfnamefont{F.}~\bibnamefont{Igloi}},
  \emph{\bibinfo{title}{preprint: cond-mat/0604126}}.

\bibitem[{\citenamefont{Brenig}(2006)}]{Brenig2006a}
\bibinfo{author}{\bibfnamefont{W.}~\bibnamefont{Brenig}},
  \bibinfo{journal}{Phys. Rev. B} \textbf{\bibinfo{volume}{73}},
  \bibinfo{pages}{104450} (\bibinfo{year}{2006}).

\bibitem[{\citenamefont{Ling et~al.}(2006)\citenamefont{Ling, Beach, and
  Sandvik}}]{Ling2006a}
\bibinfo{author}{\bibfnamefont{W.}~\bibnamefont{Ling}},
  \bibinfo{author}{\bibfnamefont{K.~S.~D.} \bibnamefont{Beach}},
  \bibnamefont{and} \bibinfo{author}{\bibfnamefont{A.~W.}
  \bibnamefont{Sandvik}}, \bibinfo{journal}{Phys. Rev. B}
  \textbf{\bibinfo{volume}{73}}, \bibinfo{pages}{014431}
  (\bibinfo{year}{2006}).

\bibitem[{\citenamefont{Sandvik}(1999)}]{Sandvik1999a}
\bibinfo{author}{\bibfnamefont{A.~W.} \bibnamefont{Sandvik}},
  \bibinfo{journal}{Phys. Rev. B} \textbf{\bibinfo{volume}{59}},
  \bibinfo{pages}{R14157} (\bibinfo{year}{1999}).

\bibitem[{\citenamefont{Sylu{\aa}sen and Sandvik}(2002)}]{sandvik02}
\bibinfo{author}{\bibfnamefont{O.~F.} \bibnamefont{Sylu{\aa}sen}}
  \bibnamefont{and} \bibinfo{author}{\bibfnamefont{A.}~\bibnamefont{Sandvik}},
  \bibinfo{journal}{Phys. Rev. E} \textbf{\bibinfo{volume}{66}},
  \bibinfo{pages}{046701} (\bibinfo{year}{2002}).

\bibitem[{\citenamefont{Sandvik}(2002{\natexlab{b}})}]{Sandvik2002b}
\bibinfo{author}{\bibfnamefont{A.~W.} \bibnamefont{Sandvik}},
  \bibinfo{journal}{Phys. Rev. B} \textbf{\bibinfo{volume}{66}},
  \bibinfo{pages}{024418} (\bibinfo{year}{2002}{\natexlab{b}}).

\bibitem[{\citenamefont{Sandvik et~al.}(1997)\citenamefont{Sandvik, Dagotto,
  and Scalapino}}]{Sandvik1997a}
\bibinfo{author}{\bibfnamefont{A.~W.} \bibnamefont{Sandvik}},
  \bibinfo{author}{\bibfnamefont{E.}~\bibnamefont{Dagotto}}, \bibnamefont{and}
  \bibinfo{author}{\bibfnamefont{D.~J.} \bibnamefont{Scalapino}},
  \bibinfo{journal}{Phys. Rev. B} \textbf{\bibinfo{volume}{56}},
  \bibinfo{pages}{11701} (\bibinfo{year}{1997}).

\bibitem[{\citenamefont{Sokol et~al.}(1994)\citenamefont{Sokol, Glenister, and
  Singh}}]{Sokol1994a}
\bibinfo{author}{\bibfnamefont{A.}~\bibnamefont{Sokol}},
  \bibinfo{author}{\bibfnamefont{R.}~\bibnamefont{Glenister}},
  \bibnamefont{and} \bibinfo{author}{\bibfnamefont{R.}~\bibnamefont{Singh}},
  \bibinfo{journal}{Phys. Rev. Lett.} \textbf{\bibinfo{volume}{72}},
  \bibinfo{pages}{1549} (\bibinfo{year}{1994}).

\bibitem[{\citenamefont{Shevchenko et~al.}(2000)\citenamefont{Shevchenko,
  Sandvik, and Sushkov}}]{Shevchenko2000a}
\bibinfo{author}{\bibfnamefont{P.~V.} \bibnamefont{Shevchenko}},
  \bibinfo{author}{\bibfnamefont{A.~W.} \bibnamefont{Sandvik}},
  \bibnamefont{and} \bibinfo{author}{\bibfnamefont{O.~P.}
  \bibnamefont{Sushkov}}, \bibinfo{journal}{Phys. Rev. B}
  \textbf{\bibinfo{volume}{61}}, \bibinfo{pages}{3475} (\bibinfo{year}{2000}).

\end{thebibliography}
\end{document}